\documentclass[aps,prl,twocolumn,showpacs]{revtex4}

\usepackage{epsf}

\begin{document}

\title{Optical absorbtion by atomically doped carbon nanotubes}

\author{I.V. Bondarev}
\altaffiliation{Also at: Institute for Nuclear Problems,
Belarusian State University, Bobruiskaya Str.11, 220050 Minsk,
Belarus}
\author{B. Vlahovic}
\affiliation{Physics Department, North Carolina Central
University, 1801 Fayetteville Str, Durham, NC 27707, USA}

\begin{abstract}
We analyze optical absorption by atomically doped carbon nanotubes
with a special focus on the frequency range close to the atomic
transition frequency.~We derive the optical absorbtion line-shape
function and, having analyzed particular achiral nanotubes of
different diameters, predict the effect of absorbtion line
splitting due to strong atom-vacuum-field coupling in
small-diameter nanotubes. We expect this effect to stimulate
relevant experimental efforts and thus to open a path to new
device applications of atomically doped carbon nanotubes in modern
nanotechnologies.
\end{abstract}

\maketitle


Carbon nanotubes (CNs) are graphene sheets rolled-up into
cylinders of approximately 1 nm in diameter.~Extensive work
carried out worldwide in recent years has revealed intriguing
physical properties of these novel molecular scale
wires~\cite{Dresselhaus,Dai}.~Nanotubes have been shown to be
useful for miniaturized electronic, mechanical, electromechanical,
chemical and scanning probe devices and as materials for
macroscopic composites~\cite{Baughman,Chou}.~Important is that
their intrinsic properties may be substantially modified in a
controllable way by doping with extrinsic impurity atoms,
molecules and
compounds~\cite{Duclaux,Shimoda,Jeong,Latil,Son}.~Recent
successful experiments on encapsulation of single atoms into
single-wall CNs~\cite{Jeong} and their intercalation into
single-wall CN bundles~\cite{Duclaux,Shimoda}, along with numerous
studies of monoatomic gas absorbtion by the CN bundles
(see~\cite{Calbi} for a review), stimulate an in-depth theoretical
analysis of near-field electrodynamical properties of atomically
doped CNs systems.~This is of both fundamental and applied
interest as it sheds light on the peculiarities of the
atom-electromagnetic-field interactions in quasi-1D dispersive and
absorbing surroundings, and, eventually, would open routes for new
challenging nanophotonics applications of atomically doped CN
systems as various sources of coherent light emitted by dopant
atoms.

Typically, there may be two qualitatively different regimes of the
interaction of an atomic state with a vacuum electromagnetic field
in the vicinity of a
CN~\cite{Bondarev02,Bondarev04,Bondarev06,Bondarev05}.~They are
the weak coupling regime and the strong coupling regime.~The
former is characterized by the monotonic exponential decay
dynamics of the upper atomic state with the decay rate altered
compared with the free-space value.~The latter is, in contrast,
strictly non-exponential and~is characterized by reversible Rabi
oscillations where the energy of the initially excited atom is
periodically exchanged between the atom and the field.~We have
recently shown that the relative density of photonic states (DOS)
near the CN effectively increases due to the presence of
additional surface photonic states coupled with CN electronic
quasiparticle excitations~\cite{Bondarev02}. This causes the
atom-vacuum-field coupling constant, which is determined by the
local (distance-dependent) photonic DOS, to be very sensitive to
the atom-CN-surface distance. At small enough distances, the
system may exhibit strong atom-field
coupling~\cite{Bondarev04,Bondarev06}, giving rise to
rearrangement ("dressing") of atomic levels by the vacuum-field
interaction with formation of atomic quasi-1D cavity
polaritons.~The problem of stability of quasi-1D polaritonic
states in atomically doped CNs has been partly addressed in
Refs.~\cite{Bondarev06,Bondarev05} by investigating the
atom-nanotube van der Waals interactions and demonstrating the
inapplicability of standard weak-coupling-based van der Waals
coupling models close to the CN surface.~In this Letter, we
analyze optical absorption properties of atomically doped CNs with
a special focus on the frequency range close to the atomic
transition frequency and predict the effect of the absorbtion line
splitting for quasi-1D atomic polaritons in small-diameter CNs.

We consider a two-level atom positioned at the point
$\mathbf{r}_{A}$ near an infinitely long achiral single-wall~CN.
The atom interacts with a quantum electromagnetic field via an
electric dipole transition of frequency $\omega_{A}$.~The atomic
dipole moment, $\mathbf{d}\!=\!d_{z}\textbf{e}_{z}$, is assumed to
be directed along the CN axis (assigned by the unit vector
$\textbf{e}_{z}$) which is chosen to be the $z$-quantization axis
of the system.~The contribution of the transverse dipole moment
orientation is suppressed because of the strong depolarization of
the transverse field in an isolated CN (the so-called dipole
antenna effect~\cite{Jorio}).~The total secondly quantized
Hamiltonian of the system is given by~\cite{Bondarev05,Bondarev06}
\begin{eqnarray}
\hat{H}&=&\int_{0}^{\infty}\!\!\!\!\!d\omega\,\hbar\omega\!\int\!d\mathbf{R}
\,\hat{f}^{\dag}(\mathbf{R},\omega)\hat{f}(\mathbf{R},\omega)+
{\hbar\tilde{\omega}_{A}\over{2}}\,\hat{\sigma}_{z}\nonumber\\
&+&\int_{0}^{\infty}\!\!\!\!\!d\omega\!\int\!d\mathbf{R}\;[\,
\mbox{g}^{(+)}(\mathbf{r}_{A},\mathbf{R},\omega)\,\hat{\sigma}^{\dag}\label{Htwolev}\\
&-&\mbox{g}^{(-)}(\mathbf{r}_{A},\mathbf{R},\omega)\,\hat{\sigma}\,]\,
\hat{f}(\mathbf{R},\omega)+\mbox{h.c.},\nonumber
\end{eqnarray}
with the three items representing the electromagnetic field
(\emph{modified} by the presence of the CN), the two-level atom
and their interaction, respectively.~The operators
$\hat{f}^{\dag}(\mathbf{R},\omega)$ and
$\hat{f}(\mathbf{R},\omega)$ are the scalar bosonic field
operators defined on the surface of the CN ($\mathbf{R}$~is the
radius-vector of an arbitrary point of the CN surface).~They play
the role of the fundamental dynamical variables of the field
subsystem and satisfy the standard bosonic commutation
relations.~The Pauli operators,
$\hat{\sigma}_{z}\!=\!|u\rangle\langle u|-|l\rangle\langle l|$,
$\hat{\sigma}\!=\!|l\rangle\langle u|$ and
$\hat{\sigma}^{\dag}\!=\!|u\rangle\langle l|$, describe the atomic
subsystem and electric dipole transitions between the two atomic
states, upper $|u\rangle$ and lower $|l\rangle$, separated by the
transition frequency $\omega_{A}$.~This (bare) frequency is
modified by the diamagnetic ($\sim\!\!\mathbf{A}^{2}$) atom-field
interaction yielding the new \emph{renormalized} transition
frequency
$\tilde{\omega}_{A}=\omega_{A}[1-2/(\hbar\omega_{A})^{2}\!
\int_{0}^{\infty}\!d\omega\!\int\!d\mathbf{R}
|\mbox{g}^{\perp}(\mathbf{r}_{A},\mathbf{R},\omega)|^{2}]$~in~the
second term of the Hamiltonian.~The dipole atom-field interaction
matrix elements
$\mbox{g}^{(\pm)}(\mathbf{r}_{A},\mathbf{R},\omega)$ are given by
$\mbox{g}^{(\pm)}\!=\!\mbox{g}^{\perp}\pm(\omega/\omega_{A})\mbox{g}^{\parallel}$
where
$\mbox{g}^{\perp(\parallel)}(\mathbf{r}_{A},\mathbf{R},\omega)\!=\!
-i(4\omega_{A}/c^{2})$ $\sqrt{\pi\hbar\omega\,\mbox{Re}\,
\sigma_{zz}(\omega)}\,\,d_{z}^{\;\,\perp(\parallel)}G_{zz}(\mathbf{r}_{A},\mathbf{R},\omega)$~with
$^{\perp(\parallel)}G_{zz}$ being the $zz$-component of the
transverse (longitudinal) Green tensor (with respect to the first
variable) of the electromagnetic subsystem and
$\sigma_{zz}(\omega)$ representing the CN surface axial
conductivity.~The matrix elements $\mbox{g}^{\perp(\parallel)}$
have the property of
$\int\!d\mathbf{R}\,|\mbox{g}^{\perp(\parallel)}(\mathbf{r}_{A},\mathbf{R},\omega)|^{2}\!=\!
(\hbar^2/2\pi)(\omega_{A}/\omega)^{2}\Gamma_{0}(\omega)
\xi^{\perp(\parallel)}(\mathbf{r}_{A},\omega)$ with
$\xi^{\perp(\parallel)}(\mathbf{r}_{A},\omega)\!=\!\mbox{Im}^{\perp(\parallel)}
G_{zz}^{\,\perp(\parallel)}(\mathbf{r}_{A},\mathbf{r}_{A},\omega)/\mbox{Im}G_{zz}^{0}(\omega)$
being the transverse (longitudinal) distance-dependent (local)
photonic DOS functions and
$\Gamma_{0}(\omega)\!=\!8\pi\omega^{2}d_{z}^{2}\,\mbox{Im}G_{zz}^{0}(\omega)/3\hbar
c^{2}$ representing the atomic spontaneous decay rate in vacuum
where $\mbox{Im}\,G_{zz}^{0}(\omega)\!=\!\omega/6\pi c$ is the
vacuum imaginary Green tensor $zz$-component.~The
Hamiltonian~(\ref{Htwolev}) involves only two standard
approximations: the electric dipole approximation and two-level
approximation.~The rotating wave approximation commonly used is
not applied, and the diamagnetic term of the atom-field
interaction is not neglected (as opposed to, e.g.,
Refs.~\cite{Bondarev02,Bondarev04,Dung}). Quantum electrodynamics
(QED) of the two-level atom close to the CN is thus described in
terms of only two intrinsic characteristics of the electromagnetic
subsystem --- the transverse and longitudinal local photonic DOS
functions.

In deriving the absorbtion/emission spectral line sha\-pe, we
follow the general quantum approach of Ref.~\cite{Heitler}.
(Obviously, the absorbtion line shape coincides with the emission
line shape if the monochromatic incident light beam is being used
in the absorbtion experiment.) When the atom is initially in the
upper state and the field subsystem is in vacuum, the
time-dependent wave function of the whole system can be written as
\begin{eqnarray}
|\psi(t)\rangle&=&C_{u}(t)\,e^{-i(\tilde{\omega}_{A}/2)t}
|u\rangle|\{0\}\rangle\label{wfunc}\\
&+&\int\!d\mathbf{R}\!\int_{0}^{\infty}\!\!\!\!\!\!d\omega\,
C_{l}(\mathbf{R},\omega,t)\,e^{-i(\omega-\tilde{\omega}_{A}/2)t}
|l\rangle|1(\mathbf{R},\omega)\rangle,\nonumber
\end{eqnarray}
where $|\{0\}\rangle$ is the vacuum state of the field subsystem,
$|\{1(\mathbf{R},\omega)\}\rangle$ is its excited state with the
field being in the single-quantum Fock state, and $C_{u}$ and
$C_{l}$ are the population probability amplitudes of the upper
state and the lower state of the whole system,
respectively.~Applying the Hamiltonian~(\ref{Htwolev}) to the wave
function~(\ref{wfunc}), one has
\begin{eqnarray} \mbox{\it\.{C}}_{u}(t)&=&-\frac{i}{\hbar}\!
\int_{0}^{\infty}\!\!\!\!\!\!d\omega\!\int\!d\mathbf{R}\,
\mbox{g}^{(+)}(\mathbf{r}_{A},\mathbf{R},\omega)\label{popampu}\\
&\times&C_{l}(\mathbf{R},\omega,t)e^{-i(\omega-\tilde{\omega}_{A})t},\nonumber
\end{eqnarray}
\vspace{-0.5cm}
\begin{equation}
\mbox{\it\.{C}}_{l}(\mathbf{R},\omega,t)\!=-\frac{i}{\hbar}\,
[\mbox{g}^{(+)}(\mathbf{r}_{A},\mathbf{R},\omega)]^{\!\ast}\,
C_{u}(t)e^{i(\omega-\tilde{\omega}_{A})t}. \label{popampl}
\end{equation}
In terms of the probability amplitudes above, the
emission/absorbtion spectral line shape $I(\omega)$ is given by
$\int\!d\mathbf{R}\,|C_{l}(\mathbf{R},\omega,t\rightarrow\infty)|^{2}$,
yielding, after the substitution of $C_{l}$ obtained by
integration [under initial conditions
$C_{l}(\mathbf{R},\omega,0)\!=\!0$, $C_{u}(0)\!=\!1$] of
Eq.~(\ref{popampl}), the result
\begin{equation}
\tilde{I}(x)=\tilde{I}_{0}(x)\left|\int_{0}^{\infty}\!\!\!\!\!\!d\tau\,C_{u}(\tau)\,
e^{i(x-\tilde{x}_{A})\tau}\!\right|^{2}\!\!, \label{Ix}
\end{equation}
where
$\tilde{I}_{0}(x)\!=\![\tilde{\Gamma}_{0}(x)/2\pi]
[(x_{A}/x)^{2}\xi^{\perp}(x)+\xi^{\parallel}(x)]$
and we have, for convenience, introduced the dimensionless
variables $\tilde{I}\!=\!2\gamma_{0}I/\hbar$,
$\tilde{\Gamma}_{0}\!=\!\hbar\Gamma_{0}/2\gamma_{0}$,
$x\!=\!\hbar\omega/2\gamma_{0}$, and $\tau\!=\!2\gamma_{0}t/\hbar$
with $\gamma_{0}\!=\!2.7$~eV being the carbon nearest neighbor
hopping integral appearing in the CN surface axial conductivity
$\sigma_{zz}$.

The upper state population probability amplitude $C_{u}(\tau)$ in
Eq.~(\ref{Ix}) is given by the Volterra integral equation
[obtained by substituting the result of the formal integration of
Eq.~(\ref{popampl}) into Eq.~(\ref{popampu})] with the kernel
determined by the local photonic DOS functions
$\xi^{\perp(\parallel)}(\mathbf{r}_{A},x)$~\cite{Bondarev04,Bondarev06}.
Thus, the numerical solution is only possible for the line shape
$\tilde{I}(x)$, strictly speaking. This, however, offers very
little physical insight into the problem of optical absorbtion by
atomically doped CNs under different atom-field coupling regimes.
Therefore, we choose a simple analytical approach valid for those
atomic transition frequencies $\tilde{x}_{A}$ which are located in
the vicinity of the resonance frequencies $x_{r}$ of the DOS
functions $\xi^{\perp(\parallel)}$. Within this approach,
$\xi^{\perp(\parallel)}(\mathbf{r}_{A},x\!\sim\!x_{r})$ are
approximated by the Lorentzians of the same
half-width-at-half-maxima $\delta x_{r}$, thus making it possible
to solve the integral equation for $C_{u}$ analytically to
obtain~\cite{Bondarev04,Bondarev06}
\begin{eqnarray}
C_{u}(\tau)\!\!\!&\approx&\!\!\!\frac{1}{2}\left(\!1+\frac{\delta
x_{r}}{\sqrt{\delta x_{r}^{2}-X^{2}}}\right)
e^{-\frac{1}{2}\left(\delta x_{r}-\sqrt{\delta x_{r}^{2}
-X^{2}}\right)\tau}\hskip0.5cm\label{Cuapp}\\
&+&\!\!\!\frac{1}{2}\left(\!1-\frac{\delta x_{r}}{\sqrt{\delta
x_{r}^{2}-X^{2}}}\right)e^{-\frac{1}{2}\left(\delta
x_{r}+\sqrt{\delta x_{r}^{2}-X^{2}}\right)\tau}\nonumber
\end{eqnarray}
with
\begin{equation}
X=\sqrt{2\delta
x_{r}\tilde{\Gamma}_{0}(\tilde{x}_{A})\xi^{\perp}(\mathbf{r}_{A},\tilde{x}_{A})}\,.
\label{X}
\end{equation}
This solution is valid for $\tilde{x}_{A}\!\approx\!x_{r}$
whatever the atom-field coupling strength is, yielding the
exponential decay of the upper state population,
$|C_{u}(\tau)|^{2}\!\approx\!\exp[-\tilde{\Gamma}(\tilde{x}_{A})\tau]$
with the rate
$\tilde{\Gamma}\!\approx\!\tilde{\Gamma}_{0}\xi^{\perp}$,~in the
weak coupling regime where $(X/\delta x_{r})^{2}\!\ll\!1$, and the
decay via damped Rabi oscillations,
$|C_{u}(\tau)|^{2}\!\approx\!\exp(-\delta
x_{r}\tau)\cos^{2}(X\tau/2)$, in the strong coupling regime where
$(X/\delta x_{r})^{2}\!\gg\!1$.

Substituting Eq.~(\ref{Cuapp}) into Eq.~(\ref{Ix}) and integrating
over~$\tau$, one arrives at the expression
\begin{equation}
\tilde{I}(x)=\tilde{I}_{0}(\tilde{x}_{A})\frac{(x-\tilde{x}_{A})^{2}+\delta
x_{r}^{2}} {\left[(x-\tilde{x}_{A})^{2}-X^{2}/4\right]^{2}+\delta
x_{r}^{2}(x-\tilde{x}_{A})^{2}} \label{Ixfin}
\end{equation}
representing the emission/absorbtion spectral line shape for
frequencies $x\!\sim\!x_{r}\!\approx\!\tilde{x}_{A}$ regardless of
the atom-field coupling strength.~The line shape is clearly seen
to be of a symmetric two-peak structure in the strong coupling
regime where $(X/\delta x_{r})^{2}\!\gg\!1$. The exact peak
positions are $x_{1,2}=\tilde{x}_{A}\pm(X/2)\sqrt{\sqrt{1+8(\delta
x_{r}/X)^{2}}-4(\delta x_{r}/X)^{2}}$, separated from each other
by $x_{1}-x_{2}\!\sim\!X$ with the Rabi frequency $X$ given by
Eq.~(\ref{X}).~In the weak coupling regime $(X/\delta
x_{r})^{2}\!\ll\!1$, and $x_{1,2}$ become complex, indicating that
there are no longer peaks at these frequencies.~As this takes
place, Eq.~(\ref{Ixfin}) is approximated with the weak coupling
condition, the fact that $x\!\sim\!\tilde{x}_{A}$ and
Eq.~(\ref{X}), to give the well-known Lorentzian
$\tilde{I}(x)\!=\!\tilde{I}_{0}(\tilde{x}_{A})/
[(x\!-\!\tilde{x}_{A})^{2}\!+\!\tilde{\Gamma}^{2}(\tilde{x}_{A})/4]$
of the half-width-at-half-maximum
$\tilde{\Gamma}(\tilde{x}_{A})/2$, peaked at $x=\tilde{x}_{A}$.

To compute the absorbtion line shapes of particular CNs, one needs
to know their transverse local photonic DOS functions
$\xi^{\perp}(\mathbf{r}_{A},\tilde{x}_{A})$ in Eqs.~(\ref{X}) and
(\ref{Ixfin}).~These are basically determined by the CN surface
axial conductivity $\sigma_{zz}$.~The latter one was calculated in
the relaxation-time approximation (relaxation time
$3\times10^{-12}$~s) at temperature 300~K.~The function
$\xi^{\perp}$ was calculated thereafter as described in
Refs.~\cite{Bondarev02,Bondarev04,Bondarev06,Bondarev05}.~The
vacuum spontaneous decay rate was estimated from the expression
$\tilde{\Gamma}_{0}(x)\!=\!\alpha^{3}x$ ($\alpha\!=\!1/137$ is the
fine-structure constant) valid for atomic systems with Coulomb
interaction~\cite{Davydov}.

\begin{figure}[t]
\epsfxsize=8.65cm\centering{\epsfbox{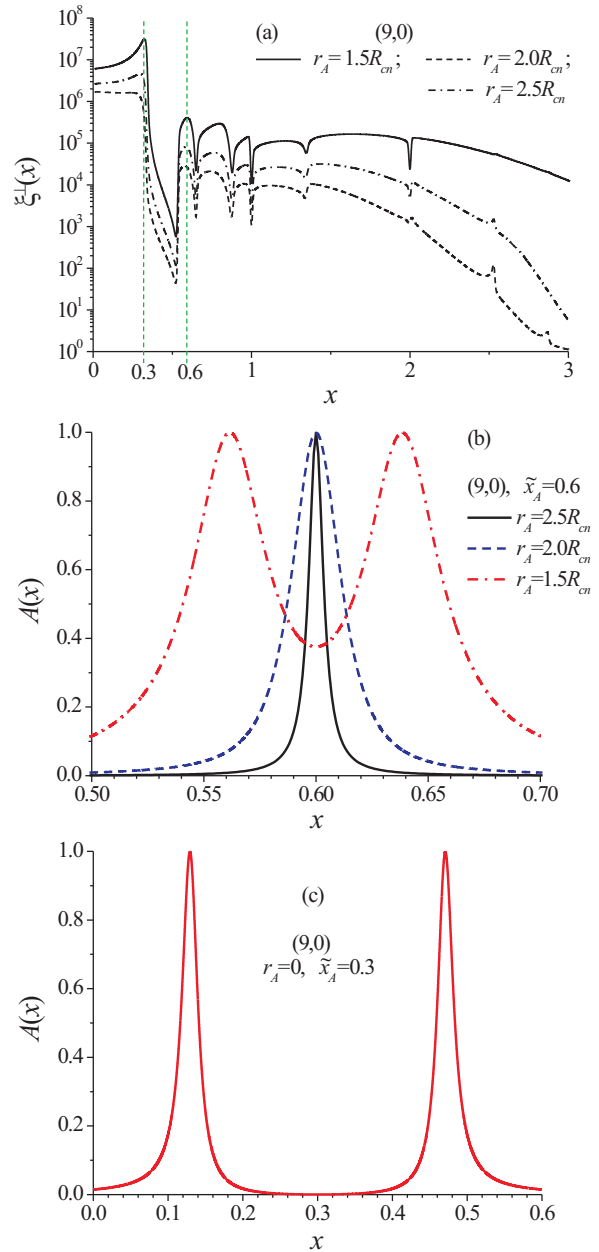}}\vskip-0.75cm
\caption{(Color online) Transverse local photonic DOS (a) and
normalized absorbtion line shapes (b) for the atom at different
distances outside the (9,0) CN. (c)~Normalized absorbtion line
shape for the atom in the center of the (9,0) CN.} \label{fig1}
\end{figure}

Figure~\ref{fig1} shows the computed $\xi^{\perp}(x)$ and
normalized optical absorption spectra
$A(x)\!=\!\tilde{I}(x)/\tilde{I}_{peak}$ for the atom close to the
(9,0) CN.~When the atom, being outside, approaches the CN surface,
the function $\xi^{\perp}$ goes up [shown in Fig.~1(a); the
vertical dashed lines indicate the DOS peak frequencies for which
$A(x)$ in Figs.~\ref{fig1}(b) and~\ref{fig1}(c) are calculated]
due to the increasing role of additional surface photonic modes of
the nanotube.~This causes the atom-vacuum-field coupling to
increase with the effect [shown in Fig.~1(b)] of first broadening
and then, when the strong atom-field coupling regime is realized,
splitting the initial Lorentzian line into two symmetric
components separated by the Rabi frequency $X\!\approx\!0.1$
corresponding to the energy
$0.1\!\times2\gamma_{0}\!=\!0.54$~eV.~In Fig.~\ref{fig1}(c) much
stronger Rabi splitting of $X\!\approx\!0.3$ is demonstrated (Rabi
energy $0.3\!\times\!2\gamma_{0}\!=\!1.62$~eV) for the atom
located in the center of the CN. This is in agreement with the
fact that, due to the nanotube curvature, the effective
interaction area between the atom and the CN surface is larger
when the atom is inside rather when it is outside the
CN~\cite{Bondarev06,Bondarev05}.

\begin{figure}[t]
\epsfxsize=8.65cm\centering{\epsfbox{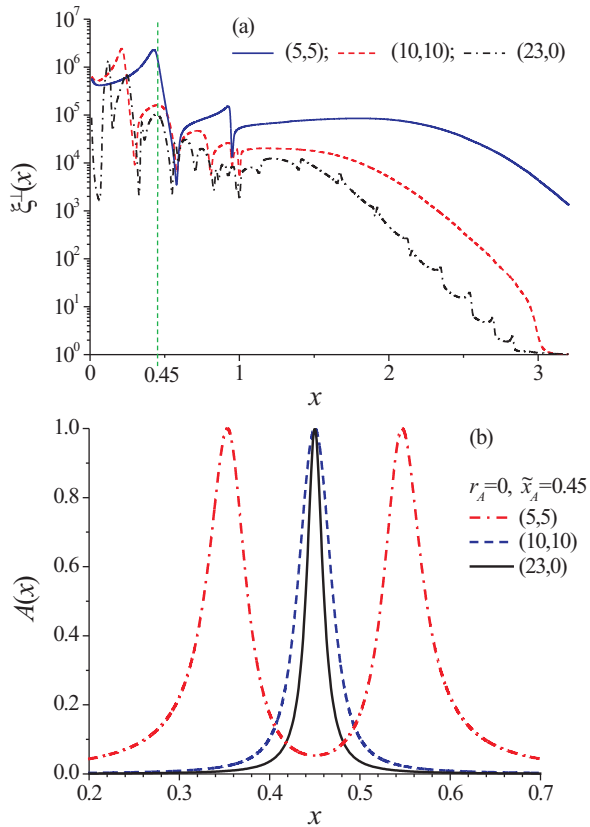}}\vskip-0.75cm
\caption{(Color online) Transverse local photonic DOS (a) and
normalized absorbtion line shapes (b) for the atom in the center
of the CNs of increasing radii.}\label{fig2}
\end{figure}

Figure~\ref{fig2}(a) presents $\xi^{\perp}(x)$ for the atom in the
center of the (5,5), (10,10) and (23,0) CNs.~It is seen to
decrease with increasing the CN radius, representing the decrease
of the atom-field coupling strength as the atom moves away from
the CN wall.~To calculate the normalized absorbtion curves
$A(x)\!=\!\tilde{I}(x)/\tilde{I}_{peak}$ in this case, we have
fixed $\tilde{x}_{A}\!=\!0.45$ [indicated by the vertical dashed
line in Fig.~\ref{fig2}(a)] because this is the approximate peak
position of $\xi^{\perp}$ for all the three CNs.~The result is
shown in Fig.~\ref{fig2}(b).~For the small radius (5,5) CN, the
absorbtion line is split into two components, indicating the
strong atom-field coupling with the Rabi splitting
$X\!\approx\!0.2$ corresponding to the energy of
$0.2\times2\gamma_{0}\!=\!1.08$~eV. The splitting is disappearing
with increasing CN radius, and the absorbtion line shape becomes
Lorentzian with the narrower widths for the larger radii CNs.

Recently, Rabi splitting $\sim\!140\!-\!400$~$\mu$eV was detected
in the photoluminescence experiments for quasi-0D excitonic
polaritons in quantum dots in semiconductor
microcavities~\cite{Reithmaier,Yoshie,Peter}.~The effect we report
in this Letter is at least 3 orders of magnitude larger.~This
comes from the fact that the typical atomic binding energies, are
at least 3 orders of magnitude larger than the typical excitonic
binding energies in solids.~In terms of the cavity-QED (see,
e.g.,~Ref.~\cite{Andreani}), the coupling constant of the
atom-cavity interaction is given in our notations by $\hbar
g=d_{z}(2\pi\hbar\tilde{\omega}_{A}/\tilde{V})^{1/2}$ with
$\tilde{V}$ being the effective volume of the field mode the atom
interacts with.~In our case $\tilde{V}\!\sim\!\pi
R_{cn}^{2}(\tilde{\lambda}_{A}/2)$, which is
$\sim\!10^2$~nm$^3\!=\!10^{-7}$~$\!\mu$m$^3$ for the CNs with
diameters $\sim\!1$~nm in the optical spectral region (as apposed
to $\sim\!10^{-1}\!\div\!10^{-2}$~$\!\mu$m$^3$ for semiconductor
microcavities in Refs.~\cite{Reithmaier,Yoshie,Peter}).~Small mode
volumes yield large atom-cavity coupling constants in
small-diameter CNs.~For example, in the situation shown in
Fig.~\ref{fig2}(a) one has $\hbar g\!\sim\!0.3$~eV for the atom in
the center of the (5,5) CN [$R_{cn}\!=\!0.34$~nm,
$d_{z}\!\sim\!er\!\sim\!e(e^{2}/\hbar\tilde{\omega}_{A})$],
whereas the "cavity" linewidth, which is related to the Purcell
factor $\xi^{\perp}(\tilde{\omega}_{A})$ via the mode quality
factor when
$\tilde{\omega}_{A}\!\approx\!\omega_{r}$~\cite{Andreani}, is
$\hbar\gamma_{c}\!=\!6\pi\hbar
c^{3}/\tilde{\omega}_{A}^{2}\xi^{\perp}(\tilde{\omega}_{A})\tilde{V}\!\sim\!0.03$,
so that the strong coupling condition $g/\gamma_{c}\gg1$ has been
well satisfied.

To summarize, we have shown that, similar to semiconductor
microcavities and photonic band-gap materials, carbon nanotubes
may qualitatively change the character of the
atom-electromagnetic-field interaction, yielding strong atom-field
coupling.~We predict the effect of the optical absorbtion line
splitting of $\sim\!1$~eV in the vicinity of the energy of the
atomic transition for quasi-1D atomic polaritons in small-diameter
atomically doped CNs.~We expect this effect to stimulate relevant
experimental efforts and thereby to open a path to novel device
applications of atomically doped~CNs.~For example, quasi-0D
excitonic polaritons in quantum dots in semiconductor
microcavities were suggested recently to be a possible way to
produce the excitonic qubit entanglement~\cite{Hughes}. Leaving
the details for a forthcoming publication, we conclude that, being
strongly coupled to the (resonator-like) cylindrical nanotube
environment, the two atomic quasi-1D polaritons can be entangled
as well, thus challenging novel applications of atomically doped
CNs in quantum communication technologies.

This work was supported by DoD and NASA via grants No
W911NF-05-1-0502 and NAG3-804.

\end{document}